# Shedding Light on Diatom Photonics by means of Digital Holography

G. Di Caprio[a], G. Coppola[a], L. De Stefano[a], M. De Stefano[b],

A. Antonucci[b,c], R. Congestri[c], and E. De Tommasi[a]

[a]*Institute for Microelectronics and Microsystems, National Council for Research, Via P. Castellino 111, I-80131 Naples, Italy*

[b]*Department of Environmental Science of the Second University of Naples, Via Vivaldi 43, I-81100 Caserta, Italy*

[c]*Laboratory of Biology of Algae (LBA), Department of Biology, University of Rome "Tor Vergata", Via della Ricerca Scientifica, I-00173 Rome, Italy* [*]

**Diatoms are among the dominant phytoplankters in the world's oceans, and their external silica investments, resembling artificial photonic crystals, are expected to play an active role in light manipulation. Digital holography allowed studying the interaction with light of *Coscinodiscus wailesii* cell wall reconstructing the light confinement inside the cell cytoplasm, condition that is hardly accessible *via* standard microscopy. The full characterization of the propagated beam, in terms of quantitative phase and intensity, removed a long-standing ambiguity about the origin of the light. The data were discussed in the light of living cell behavior in response to their environment.**

*Short Title*:  G. Di Caprio et al*., Shedding Light on Diatom Photonics.*

*Manuscript Keywords: Digital Holography, Marine Biology, Diatoms, Microscopy*

Corresponding author: Dr. Giuseppe Di Caprio, e-mail: *dicaprio@rowland.harvard.edu*
Rowland Institute at Harvard, 100 Land Boulevard, 02142 Cambridge. Tel: +1 (617) 497 4714; Fax: +1 (617) 497 - 4627

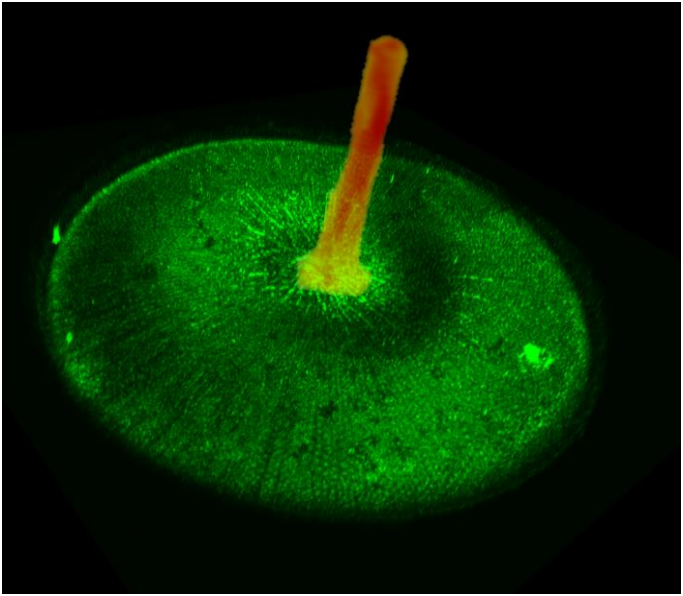

Abstract figure - Overlay of a cell reconstructed by Confocal Laser Scanning Microscopy and the 3D representation of light confinement along z-axis obtained via Digital Holography.

The development of scanning and transmission electron microscopy allowed to discover and to study in great detail the presence of periodic or quasi-periodic structures, at micro and sub-micron scale, as constituent parts of various plants and animals (1). These structures, incorporated as deformed or quasi-ordered multilayer arrangements within the morphology of algae, plants and insects, mainly act as photonic crystals, manipulating light by means of interferometric and diffractive effects. This can result in the presence of structural bright colors in a number of butterflies and beetles which are not determined by the presence of a pigment, but are the consequence of photonic band-gaps, allowing reflection only in narrow spectral ranges (2, 3). Other examples of photonic structures in biology include the exocuticle of the beetle *Chrysina gloriosa,* which is able to selectively reflect left circular polarized light (4), the microlense arrays of dorsal arm plates in the brittlestar *Ophiocoma wendtii* (1), and the wide anti-reflective surface, based on sub-micron nipple arrays, covering the eye of several moth species (5). The realization of novel micro- and nano-optical devices aimed at manipulating light by means of selective reflection, transmission, strong polarization control and spectral filtering can benefit of bioinspiration from these photonic structures optimized by selective pressure during life's history on earth. This actually constitutes the basis of the so-called bio-mimicry or bio-mimetics (5).

One of the most versatile photonic architecture recognized in nature is represented by the diatom cell walls. Diatoms are unicellular microalgae provided with external silica investments, the frustules (6). Frustules are constituted by two halves (the valves) interconnected by a lateral girdle. Valves and girdles exhibit hierarchical, complex and quasi-ordered patterns of pores (*areolae*) whose dimensions can range from nanometer to micrometer scale (depending on species and position in the cell wall structure). Several studies suggest that these patterns, not only serving a protective function against mechanical stress (shear forces, fluid properties), animal grazing and other noxious agents, confer to frustules an active role in manipulation and exploitation of light, ultimately affecting photosynthetic performance (7-9). The impressive similarity of such structures with artificial photonic crystals enforces this hypothesis (7). The study of optical and photonic properties of diatom frustules is thus attractive as understanding of the evolutionary design principles and advantage of pore patterns is highly prospect to inspire novel developments in nanotechnology, optical sensing and biosensing (10).

Diatoms are among the most successful eukaryotic micro-organisms in aquatic environments, they are nearly ubiquitous, adapted to the many different aquatic environments. These unicellular or colonial microalgae (1-1000 μm) are of immense ecological significance, contributing one quarter of global primary productivity and hence to food webs and carbon cycle. Diatoms are premier organisms involved in biosilicification, both in terms of the amount of silica they process and in the variety of silicified structure they can make. About $10^5$ diatom species have been described up to now, mainly classified for their shape, morphometry and pore patterns. Two main groups can be distinguished, based on valve symmetry and areolar design: centrics are characterized by radial symmetry of frustules while pennate diatoms have elongated and bilaterally symmetric valves. Among radial centrics, *Coscinodiscus wailesii* Gran & Angst 1931 has been the object of several studies focusing on the optical and photonic properties of frustules. Early studies have demonstrated the ability of a single valve to confine coherent radiation coming from a diode laser in a spot a few micron wide (11). Further experiments and numerical simulations showed that this confinement also occurred with non-coherent radiation, probably due to in-phase superposition of the diffracted waves arising from pores (12).

In this work, a novel approach based on Digital Holography (DH) allowed to further investigate on *C. wailesii* photonic properties namely its valve interaction with coherent light. DH provides detection of three-dimensional features of biological microstructures (13-15), and enables to quantitatively retrieve, in far field region, the amplitude and phase of the wavefront interacting with the structures themselves (16, 17). The usual scanning analysis of the field transmitted by a diatom valve (11) suffers from two serious constraints: the resolution in the direction of propagation of the light (namely z-direction) is limited by the mechanical scanning pitch of the microscope objective in use and the case of light propagating in cytoplasm is not experimentally accessible. Both the limitations have been successfully overcome in this work by the use of DH. In particular, the resolution of the retrieved field is sub-wavelength along the z-direction. Moreover DH permitted to clarify the role that the pores and the valve edges play in light manipulation, discriminating the two contributions.

## Results and Discussion

**Morphological and ultrastructural characterization of *Coscinodiscus wailesii* frustule.** In order to obtain an insight into valve ultrastructure, Scanning Electron Microscopy (SEM, JEOL JSM-6060LV) images have been acquired (Fig. 1a-d). Valve structure micrographs revealed the presence of two silica layers or plates (loculate valves): the external one is perforated by a complex array of hollow pores of about 250 nm in diameter with a lattice constant around 500 nm while the internal layer exhibits hexagonally spaced pores whose diameter ranges between 1.2 and 1.5 µm and the lattice constant is about 2 µm. Atomic Force Microscopy (AFM, XE-70 Park's) provided a more resolved characterization of the valve sub-micrometric features. Detailed topographies of the valve inner plate were obtained while ultrastructure of the outer plate is detectable through its pores (Fig. 1e-h).

**Amplitude and phase retrieval of the wavefront in air.** The interference pattern of the light transmitted by *C. wailesii* valve and recombined with reference beam has been acquired by a CCD camera and the corresponding intensity profile is shown in Fig. 2a. Subsequently, the reconstructed optical wave field has been propagated at different distances by means of a proper numerical algorithm (for analytical details see equation (3) in *Materials and methods*) along the forward direction (i.e., the z-axis). The amplitude reconstruction at z=200 µm is reported in Fig. 2b. Light confinement at the valve center, described elsewhere using standard optical microscopy (11, 12), is clearly visible, while reconstruction at z=400 µm (Fig. 2c) shows a spread of the central light spot and, simultaneously, the appearance of light rings due to the diffraction from valve edges. In order to hide external light and to highlight the studied process, the three intensity profiles have been filtered by means of a binary mask. As already deduced from Wide-Angle Beam-Propagation-Method numerical simulations (12), the narrow light confinement along the optical axis (z direction) is mostly due to coherent superposition of the diffractive contributions coming from valve pores.

Past research allowed us to measure the spatial distribution of light transmitted by a single valve recording series of images acquired at different positions along the optical axis by means of a microscope objective (11, 12). This method suffered from the limitation due to the mechanical scanning pitch, within which the amplitude information was averaged. In present work, thanks to the interferometric nature of the holographic acquisition, we got a resolution along the z direction of $\sim\lambda/20$, thus providing a more refined characterization of the z-propagation as compared to the above measurements. A numerical flat illumination field, whose diameter is much larger than the dimension of the diatom, has been multiplied to the reconstructed optical field in the diatom plane and numerically propagated. The plots of the intensity values have been obtained along a valve diameter and 3D reconstruction of light confinement along the z axis is shown (Fig. 2d). XZ- and YZ- cross sections of intensity reconstructions were also obtained (Fig. 2 e and f). X and Y intervals are limited to the areas in which the diatom is present. The confinement effect in the region around z=200 µm is visible, as well as the diffraction contributions from the valve edges.

The unique capability of DH to manage the phase information allows adding information to the characterization of the light transmission by a *C. wailesii* valve. The optical field has been numerically reconstructed at several, different z positions between 0 and 600 µm and the corresponding phase profiles plotted in a range of 20 μm around the central intensity peak, both in the X- and the Y-directions (Fig. 2g and 2h respectively). Besides the expected irregularities of the plotted reconstructions, as the object tested is far from being perfectly symmetric, the phase profile behavior appears similar to that of a light beam around the waist position (27). In fact, an opposite concavity of the phase, around a region where profiles appear flatter, is shown, the flatness region is compatible with the high confinement region in Fig. 2e and 2f.

**Amplitude and phase retrieval of the wavefront in cytoplasm.** The potentiality of the operator algebra, described in *Materials and methods*, can be fully exploited substituting the refractive index of air with that of the medium of interest in the expression of the wavenumber k in equation (3). In our case it is of great interest to understand how light is manipulated by the valve and then distributed inside the living cell, an information which can hardly be obtained by means of traditional optical microscopy. In a first approximation this knowledge can be straightforwardly obtained by substituting in equation (3) the refractive index of air (n=1) with that of cytoplasm (n=1.35). The results in terms of intensity maps at different positions along the optical axis, intensity distributions in the XZ and YZ planes and spatial evolution of phase along the direction of propagation are shown in Fig. 3. It is clear how light confinement takes place at a lower distance from the valve (the maximum of transmitted light takes place at about z=130 μm in case of cytoplasm versus z=200 μm in case of air, however most of transmitted light is confined starting from z=100 μm in case of cytoplasm, versus z=150 μm in case of air). As the whole frustule is in principle constructed like a box, consisting of two valves joined together by a girdle (in which protoplasts are enclosed), during cell life cycles, it appears as a barrel-shaped shell whose length is comparable with valve diameter. Thus, during light propagation in the cytoplasm, radiation is mostly confined inside the frustule, making the collected light available to the numerous, discoid plastids or chloroplasts (Fig. 4). The three-dimensional spatial distribution of plastids inside a living diatom cell can be efficiently detected by means of confocal laser scanning microscopy (27) (see Fig. 4c). The autofluorescence of diatom chloroplasts, characterized by a red emission and the intrinsic photoluminescence of the frustule in the green portion of the spectrum, mainly attributable to surface defects of the hydrated amorphous silica nanostructure (29) and already exploited as transducing mechanism in gas-sensing experiments (30, 31), are indeed excited by illuminating the *C. wailesii* culture sample with a diode laser (635 nm) and an Argon ion laser (488 nm), respectively. The relationship between relocation of plastids under given illumination, photosynthetic performance, light availability and its spectral content, is not straightforward (9) and accurate *in vivo* testing of *C. wailesii* grown under exposure to different monochromatic lights is needed. Still some reasonable assumptions can be drawn starting from empirical evidences collected so far. The frustule has revealed to represent a significant defense against environmental constraints, both as an effective filter against external agents and by virtue of its solid mechanical resistance at the diatom scale.

This resistance, mainly due to dispersion of mechanical stresses through the silica porous matrix (32), allows the frustule to withstand pressures of the order of hundreds of tons per square meter (33). On the other hand, the presence of the frustule itself determines a lower exposition of diatom cells to sunlight. In this scenario, the regular areolar pattern of diatom frustules could represent an elegant mode generated under selective pressure, of efficient light harvesting by means of the diffractive mechanisms described here. Furthermore, it is known that, under intense illumination conditions, the diatom chloroplasts migrate away from the cell walls moving towards the center of the intracellular space, where we observed the light confinement effect (see Fig. 4b). In particular, first studies on the centric diatom *Pleurosira laevis* demonstrated the photosynthetic organelle migration to the cortical cytoplasm under weak white light illumination, while, under intense irradiation, chloroplasts migrated towards the nucleus (34). Further research allowed to integrate this information with the plastids spectral response in *P. laevis*, irradiation with blue light induced migration toward the nuclear cytoplasm, while green light determined movement toward the cortical one (35). Noteworthy maxima of photosynthesis action spectra, i.e. the efficiency of the photosynthetic process, are found, for most diatom species, in the blue and in the red regions of the visible (36). Our previous studies showed that, as long as visible wavelengths are concerned, light confinement along the optical axis of the frustule always takes place, even if the precise position in z slightly varies with wavelength due to dispersion of silica refractive index (12). Thus, as long as light is available in the external environment, light is always accessible also along the diatom cell axis, despite the presence of silica walls and in virtue of the confinement effect induced by pore patterns. In this context, chloroplast migration toward nuclear cytoplasm, which is most likely mediated by $Ca^{2+}$ permeable channels (35), seems to occur only for those wavelengths that guarantee the maximum efficiency of photosynthesis and for sufficiently high intensities. Finally, if we look at transverse intensity distribution inside and outside the surface delimited by the valve (see Fig. 5), we observe that, even in condition of maximum light confinement (z=130 μm for cytoplasm), the light intensity inside the diatom is slightly lower than outside and, as a consequence, light confinement effect does not determine inhibition of photosynthesis (36), when light in the environment is below the photo-inhibition threshold.

Furthermore, the strong dependence of the spatial location of light confinement by the refractive index of the surrounding environment, makes this effect an ideal transduction mechanism in a possible, bio-inspired, label-free micro-optical sensing scheme (37).

**Discriminating the different contributions to diffraction.** DH can be efficiently exploited in order to distinguish and separate the diffraction contribution of the valve edges from that due to pore pattern. A disk-shaped obstacle, of the same dimensions and contour as the valve, has been used for application of the propagation algorithm previously described, typical diffraction rings coming from the borders of the disk are clearly visible in different positions along the optical axis (Fig. 6a). Indeed, in Fresnel approximation, the diffraction pattern behind a circular disk is characterized by a concentric ring structure and a bright central spot, known as Poisson's spot (38). Poisson's spot takes place at a distance much greater than the one where light confinement occurs. In fact, a spot is not visible over a distance of hundreds of microns from the z=0 origin (Fig. 6a). On the other hand, when the hologram of the diatom valve is illuminated by a gaussian distribution of light with a standard deviation involving only the pores and not the valve edges (Fig. 6b, FWHM about 45 $\mu$m for the incoming distribution of intensity), and the propagation algorithm is applied, light confinement is seen to take place at the expected distance along the optical axis. This confirms, on the basis of an experimentally acquired hologram, and after numerical simulations (reported in detail elsewhere 12), that the origin of the confinement effect in a *C. wailesii* valve is only due to diffraction of pores with no contribution from valve edges.

**Conclusions**

Manipulation of light by a single valve of the marine centric diatom *C. wailesii*, has been thoroughly investigated by means of digital holography. The proper operator algebra allowed to reconstruct the optical field after the interaction with the diatom valve at ultrastructural level and at different positions along the optical axis. By virtue of the interferometric resolution guaranteed by the holographic technique, light confinement effect has been detected with unprecedented refinement along the optical axis and the evolution of the phase of the optical field has been retrieved for the first time. The same analysis has been performed in cytoplasm environment and a strong reduction of the distance at which confinement occurs has been observed. The diffractive origin of the confinement effect has been also clarified by efficiently separating the diffraction contributions of the different valve structural features. Further *in vivo* studies of *C. wailesii* cultures exposed to different illumination conditions are foreseen, in order to clarify the complex relationship between optical field distribution inside the living cells and the spatial relocation of chloroplasts under different spectral solicitations.

**Materials and Methods**

**Sample preparation.** *C. wailesii* inocula have been purchased from the Culture Collection of Algae and Protozoa of the Scottish Association for Marine Science, UK. Enriched cultures were maintained in Guillard "f/2" medium with silica addition [6] at a temperature of 18 °C and illuminated by white light fluorescent tubes (PHILIPS HPL-N150W), with an irradiance of 30-50 $\mu$mol photons m$^{-2}$ s$^{-1}$ and 14:10 LD cycle. Frustules of cultured cells were cleaned by strong acid solutions and heating according to von Stoch's method [17], which provided organic matter removal and random separation of frustule elements. The obtained, sparse valves, characterized by an average diameter of about 190 $\mu$m, were deposited on a glass slide for the subsequent optical and holographic analyses.

**Holograms acquisition.** A He-Ne laser beam ($\lambda$ = 633 nm) is splitted into a reference and an object beam by a cube-polarizing beam splitter. Both the object and the reference beams are filtered and expanded. A microscope objective (magnification: 10X; NA: 0.22) has been used to collect the object beam. Object beam and reference beam are then recombined by a second beam-splitter onto a charged-coupled-device (CCD) detector (1392 × 1040 pixel array; pixel size $\Delta x = \Delta y$ = 4.7 μm), which acquires the formed interference pattern. A $\lambda$/2 waveplate rotates the polarization state of the object beam in order to optimize the fringe contrast. The setup is shown in Fig. S1.

**Theoretical background.** The digital hologram reconstruction procedure allows to retrieve a discrete version of the complex optical wavefront diffused by the object under test. The complex field of the object beam (see Methods for the details of the experimental set-up) is numerically reconstructed from the frequency spectrum of the acquired hologram. In order to obtain the spatial separation of three diffraction terms without overlapping, a configuration with a small angle between the reference beam and the object beam (*off-axis configuration*) is adopted (18). Thus, the first diffraction order can be separated from the whole spatial frequency spectrum with a band-pass filter and shifted to the origin of the plane, obtaining the spectrum of the object field (defined as $O(x,y) = |O(x,y)|e^{i\varphi(x,y)}$, with $|O(x,y)|$ and $\varphi(x,y)$ amplitude and phase, respectively, and x and y cartesian coordinates defining the plane of acquisition of the hologram), except for a constant (19). As the whole field is known, it is possible to reconstruct the optical wavefront at different distances from the plane of acquisition applying the Fourier formulation of the Fresnel-Kirchhoff diffraction formula (20). An interesting approach for the analysis of the propagation problem by means of the operator algebra, compatible with our propagation range, has been proposed by Sahimir (21). Fresnel diffraction is described by replacing the Fresnel-Kirchhoff integral, the lens transfer factor, and other operations by operators. The resulting operator algebra leads to the description of Fourier optics in a simple and compact way, bypassing the cumbersome integral calculus. The detail of the formalism can be found in (22). By means of this approach the propagated field $O_{prop}(x,y)$ as a function of the initial field $O(x,y)$ can be rewritten as

$$O_{prop}(\xi,\eta) = \exp(ikd)\left\{\mathfrak{I}^{-1}\left[exp\left(-\frac{ikd\lambda^2}{2}(v^2+\mu^2)\right)\right]\cdot\mathfrak{I}(O(x,y))\right\} (1)$$

being $\mathfrak{I}[f(x)]$ and $\mathfrak{I}^{-1}[f(x)]$ the Fourier transform and anti-transform, respectively, of the function $f(x)$, $k = \frac{2\pi n}{\lambda}$ (with $n$ refractive index of the medium), $n$ and $m$ spatial frequencies defined as $v = \frac{\xi}{\lambda d}$ and $\mu = \frac{\eta}{\lambda d}$, and $d$ the reconstruction distance. For digital reconstruction Eq. (1) is implemented in a discrete form (23)

$$O_{prop}(m,n) = \exp(ikd)\left\{\mathfrak{I}_D^{-1}\left[exp\left(-\frac{ikd\lambda^2}{2N^2\Delta^2}(U^2+V^2)\right)\right]\cdot\mathfrak{I}_D(O(h,j))\right\} (2)$$

where N is the number of pixels in both directions, Δ is the sampling distance (i.e. the pixel dimension), *m*, *n*, *U*, *V*, *h* and *j* are integer numbers varying from 0 to *N*-1. The discretized Fourier transform is defined as

$$\Im_D\{g(h,j)\} = \frac{1}{N}\sum_{j,l=0}^{N-1} exp\left[-\frac{2\pi i}{N}(mh+nj)\right] g(h,j) \quad (3)$$

Intensity and phase distributions can be retrieved starting from the propagated field according to the following relations:

$$I_{prop}(m,n) = |O_{prop}(m,n)|^2 \quad (4) \qquad \varphi_{prop}(m,n) = arctan\frac{Im[O_{prop}(m,n)]}{Re[O_{prop}(m,n)]} \quad (5)$$

Since the phase distribution is obtained by a numerical evaluation of the *arctan* function, the values of the reconstructed phase are restricted in the interval [–π, π], i.e., the phase distribution is wrapped into this range. In order to resolve possible ambiguities arising from thickness differences greater than *λ*/2, phase-unwrapping methods have to be generally applied (22, 23). The possibility offered by DH to manage the phase of the reconstructed image allows to remove and/or compensate the unwanted wavefront variations (such as optical aberrations, slide deformations etc.) (24, 25). In this paper, a double exposure technique (26) has been used. The first exposure is made on the object under investigation, whereas the second one is made on a flat reference surface in proximity of the object. This second hologram contains information about all the aberrations introduced by the optical components including the defocusing due to the microscope objective. By numerically manipulating the two holograms, it is possible to compensate for these aberrations.

**Acknowledgements**

The authors would like to thank Dr. Elena Romano, from the Centre of Avanced Microscopy, Department of Biology, University of Rome " Tor Vergata", for her skillful assistance in the use of the CLSM facility.

This work was partially supported by MIUR (Italian Ministry for Education, University and Research) through the project "Photonic and Micromechanical Properties of Diatoms", in the framework of FIRB (Investment Fund for Basic Research) – "Future in Research" 2008.

**Figure Legends**

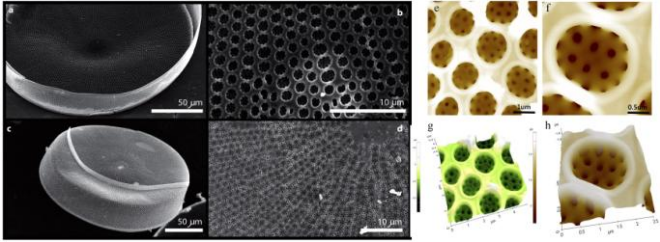

**Figure 1 - Diatom valve imaging by scanning microscopy:** SEM micrographs of valves. Outer view of valve face (a) and a detail of the external pore pattern (b); valve inner view (c) and inner pore arrangement (d). AFM images of *C. wailesii* valve portions. Topographies over 5×5 and 2.5×2.5 µm$^2$ areas are visible in (e) and (f) respectively. 3D reconstructions of the same areas of interest are shown in (g, h). Ultrastructure of the outer plate is clearly evidenced through pores of the inner silica layer.

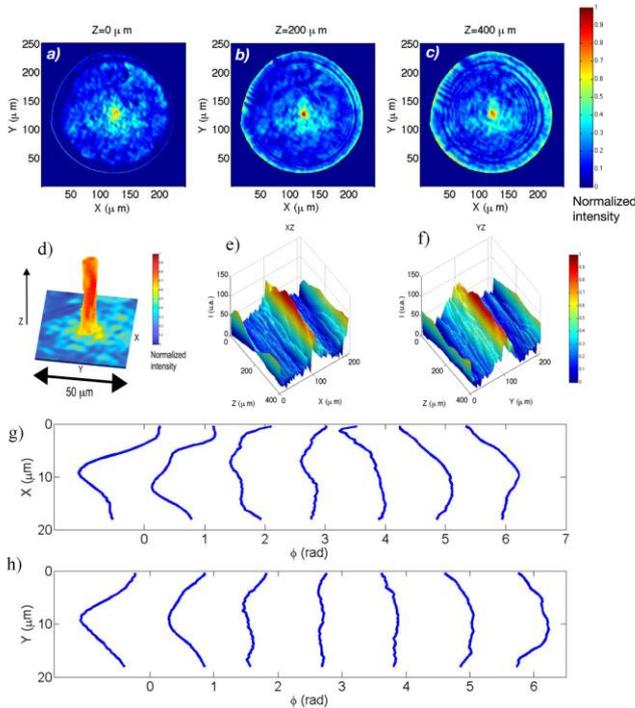

**Figure 2 - Reconstruction of the light transmitted by a diatom valve in air:** Intensity map reconstructed at three different distances: Z=0 µm (a), Z=200 µm (b), Z=400 µm (c). (d) 3D representation of light confinement along z axis. Intensity maps have been evaluated at different z values (over a full distance of 400 µm) and lined up. Intensity values below 0.5 have been removed for clearness. The number of z values has been optimized for the same reason. XZ (e) and YZ (f) distribution of the reconstructed transmitted light intensity. Diffraction from the edges of the valve is also visible at the limits of X and Y intervals. Phase profiles at seven different Z positions (Z= 0, 100, ... 600 µm) along the X (g) and Y (h) directions.

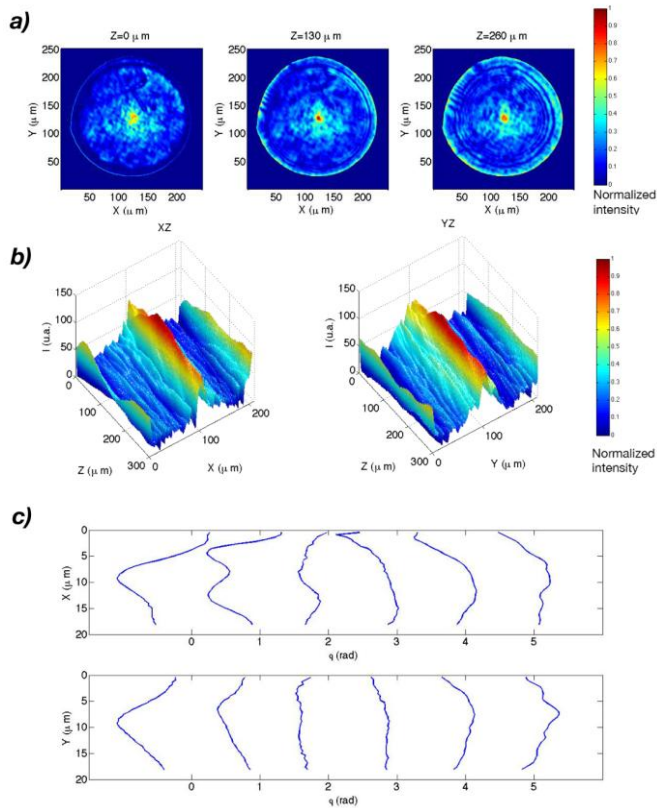

**Figure 3 - Reconstruction of the light transmitted by a diatom valve in cytoplasm:** Intensity maps at different distances along the z axis (a), intensity distributions along the XZ and XY planes (b) and evolution of the phase along the X and Y directions at six different Z positions (c) in the case of propagation in cytoplasm. The confinement effect takes place at a lower distance respect to propagation in air.

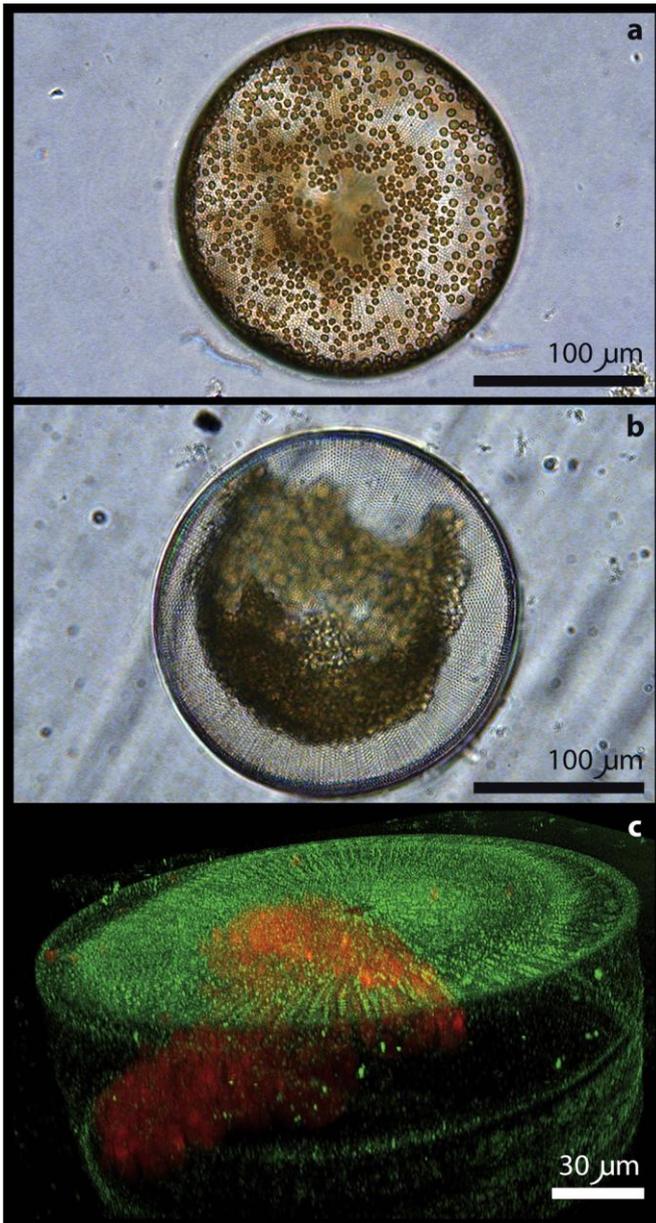

**Figure 4 - Chloroplast spatial distribution in living *C. wailesii* cell.** Light micrograph of a single cell with chloroplasts evenly distributed in xy plane **(a)**. Chloroplast agglomeration after exposure to intense white irradiation **(b)**. Three-dimensional reconstruction of a cell obtained by CLSM (Olympus Fluoview 1000): red emission comes from chloroplast autofluorescence, while green emission is due to silica frustule photoluminescence. Chloroplast position rearrangement, induced by laser excitation, is visible in volume rendering **(c)**.

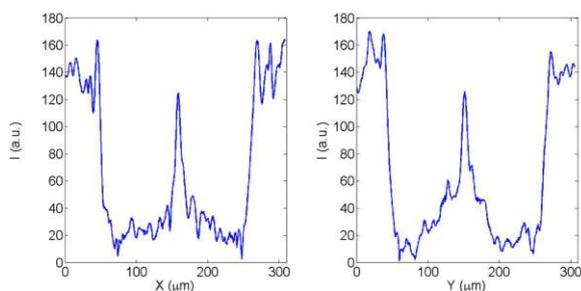

**Figure 5 - Reconstructed intensity distributions along X and Y directions for z=130 microns in cytoplasm**: The discontinuity in intensity is to be ascribed to the transit between the inner and the outer space delimited by the valve.

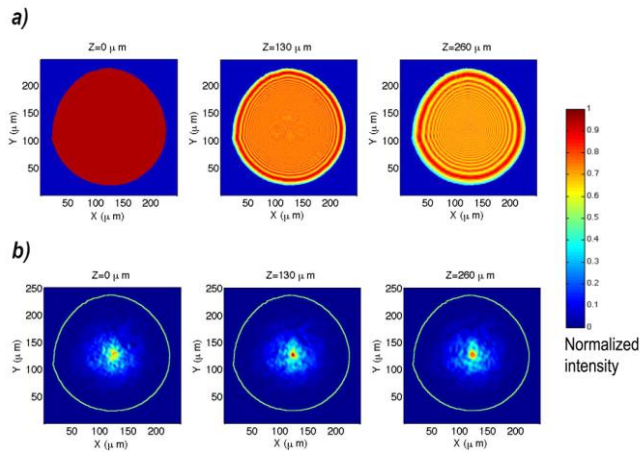

**Figure 6 – Separating the role of the edge from the nanopatterned structure:** Diffraction from the edges of a disk with the same dimensions and contour of the diatom valve at three different z positions along the optical axis **(a)**: typical ring structure associated to diffraction from a disk are clearly visible, but no central spot takes place within a distance of hundreds of microns; intensity maps in case of gaussian illumination (standard deviation of about 45 μm such as not to involve diatom valve edges) **(b)**: light confinement takes place at the same distance as in the case of rectangular illumination involving the entire valve. Both cases refer to propagation of light in cytoplasm.

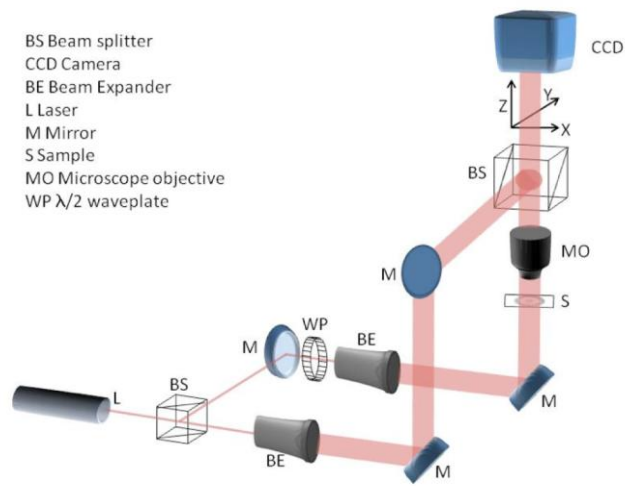

S1: **Experimental set-up for the acquisition of holograms from single diatom valves.**

**Giuseppe Di Caprio** graduated cum laude in Physics and received his PhD at the University of Naples "Federico II. He spent 7 months as invited researcher at the *Université Libre de Bruxelles* and he is currently post-doctoral researcher at the Harvard University. As a graduate student he worked on the application of DHM for the study of biological samples and his research activity is now dedicated to the development of a new generation of optical diagnostic devices.

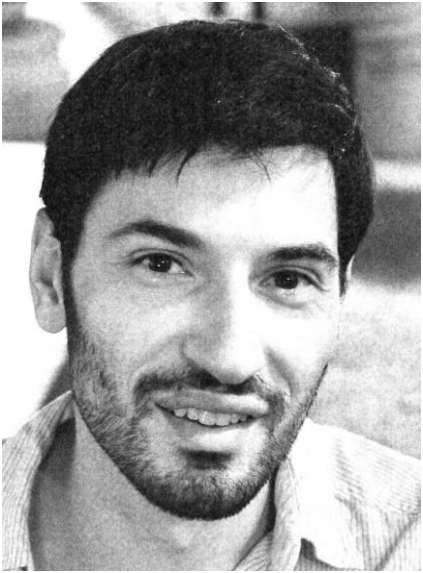

**Giuseppe Coppola** received the M.Sc degree in electronic engineering and the Ph. D. in 1997 and 2001, respectively, both from the University of Napoli "Federico II," Italy. In 2001, he spent six months as Visiting Scientist at the Technical University of Delft. Since 2002, he has been a Researcher at the Institute for Microelectronics and Microsystems of the CNR. He is now pursuing his interests in the design and fabrication of optoelectronic devices, and their characterization with DH.

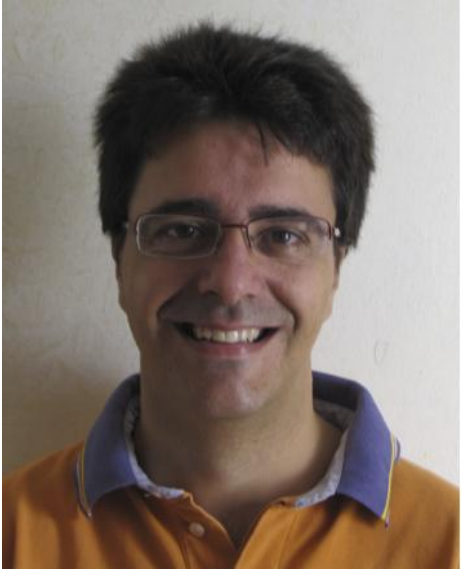

**Luca De Stefano** graduated cum laude in Physics at University of Naples "Federico II" in 1992 and received the Ph.D. in Physics in 1996. He is senior researcher at the Institute for Microelectronic and Microsystems of National Research Council in Naples, in the fields of biophotonics and optical microsystems for biochemical sensing. He presented his work to more than 150 national and international conferences and he is author of more than 90 scientific articles published on peer reviewed journals.

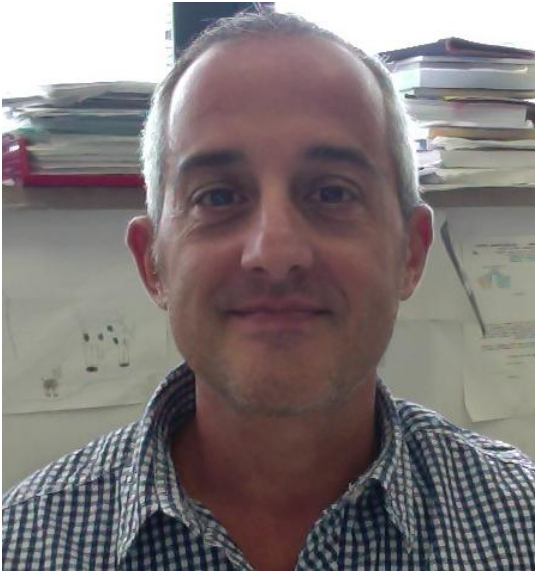

**Mario De Stefano** is a Researcher at the Department of Environmental Science of the Second University of Naples. He obtained his Ph.D. from the Stazione Zoologica "A. Dohrn" of Naples and Messina University in 2002. He is involved in basic researches on the life history of Mediterranean, Polar and tropical diatom communities, and in applied researches on the potential use of diatom for nanotechnological application in the field of optoelectronics, nanomechanics, and biosensoring.

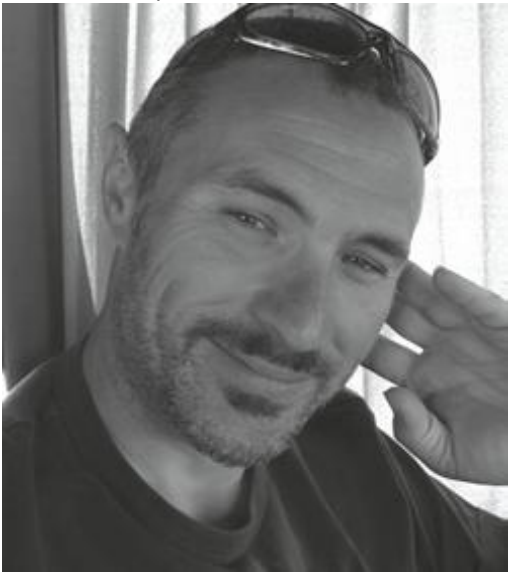

**Alessandra Antonucci** obtained her B.Sc. in Molecular and Industrial Biotechnology at "Federico II" University, in Naples and her M.Sc. in Environmental and Industrial Biotechnology at "La Sapienza" University in Rome. In 2010 she did a traineeship at the University of Rome "Tor Vergata" centred on the study of microalgae ecophysiology and their environmental applications. Currently she joins a multidisciplinary research group involved in a national research project focused on photonic and mechanical properties of diatom frustules.

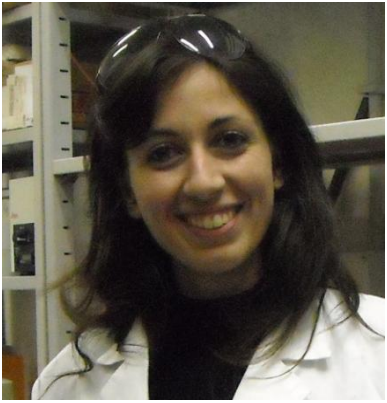

**Roberta Congestri** is a researcher in the Department of Biology, University of Rome "Tor Vergata" where she leads a research group working on biology and biotechnology of microalgae. She received her PhD in Biology of the Algae in 2002. Her research focuses on ultrastructure, taxonomy and ecophysiology of microalgae and cyanobacteria in aquatic biofilms and marine phytoplankton, with applications in diatom photonics, nanotechnology and biodesign, algal biomass production and wastewater remediation.

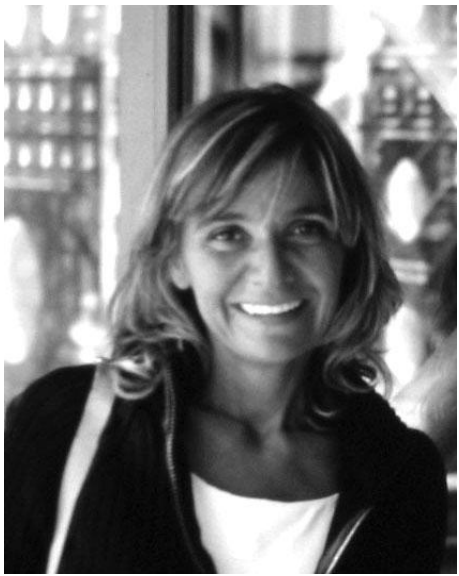

**Edoardo De Tommasi** received the Laurea degree in Physics from the University of Naples in 2002 and the PhD degree in "Novel Physical Methodologies for Ecological Research" from the Second University of Naples in 2007. Since then he has been employed at Institute for Microelectronics and Microsystems of the CNR, where his activities were focused on the development of optical biosensors based on porous silicon technology and on the study of the photonic properties of marine diatoms.

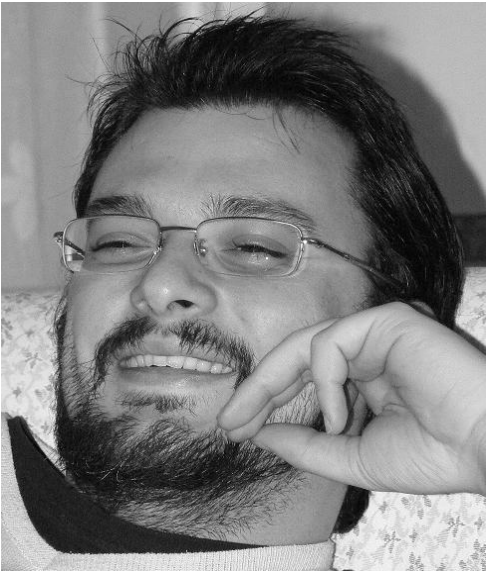